\documentclass[%
reprint,
superscriptaddress,
amsmath,amssymb,
aps,
longbibliography,
pra
floatfix,
twocolumn
]{revtex4-2}

\newcommand{\bra}[1]{\langle #1\rvert}
\newcommand{\ket}[1]{\lvert #1\rangle}

\usepackage{comment}

\usepackage{tabto}
\usepackage{graphicx,float}
\usepackage{dcolumn}
\usepackage{bm}
\usepackage{hyperref}
\usepackage{bigints}
\usepackage{amssymb}

\usepackage{tikz}
\usetikzlibrary{calc}

\usepackage{amsmath}
\usepackage{upgreek}
\usepackage{physics}
\usepackage{isomath}
\usepackage{placeins}
\usepackage{comment}
\usepackage{mathrsfs}

\begin{document}


    \title{Dynamical phases of a BEC in a bad optical cavity at optomechanical resonance}
    \author{Gage W. Harmon}
    \affiliation{Theoretische Physik, Saarland University, Campus E2.6, 66123 Saarbrücken, Germany}
    \author{Giovanna Morigi}
    \affiliation{Theoretische Physik, Saarland University, Campus E2.6, 66123 Saarbrücken, Germany}
    \author{Simon B. J\"ager}
    \affiliation{Physics Department and Research Center OPTIMAS, Technische Universität Kaiserslautern, D-67663, Kaiserslautern, Germany}
    \date{\today}

    \pacs{Valid PACS appear here}

    \begin{abstract}
        We study the emergence of dynamical phases of a Bose-Einstein condensate that is optomechanically coupled to a dissipative cavity mode and transversally driven by a laser.  We focus on the regime close to the optomechanical resonance, where the atoms'
      refractive index shifts the cavity into resonance, assuming fast cavity relaxation. We derive an effective model for the atomic motion, where the cavity degrees of freedom are eliminated using perturbation theory in the atom-cavity coupling and benchmark its predictions using numerical simulations based on the full model. Away from the optomechanical resonance, perturbation theory in the lowest order (adiabatic elimination) reliably describes the dynamics and predicts chaotic phases with unstable oscillations. Interestingly, the dynamics close to the optomechanical resonance are qualitatively captured only by including the corrections to next order (non-adiabatic corrections). In this regime we find limit cycle phases that describe stable oscillations of the density with a well defined frequency. We further show that such limit cycle solutions are metastable configurations of the adiabatic model. Our work sheds light on the mechanisms that are required to observe dynamical phases and predict their existence in atom-cavity systems where a substantial timescale separation is present.
    \end{abstract}

    {
        \let\clearpage\relax
        \maketitle
    }

    \section{Introduction}

Ultracold atomic gases interacting with an optical cavity have become a very versatile platform to study out-of-equilibrium physics. In this setup the cavity field mediates long-range interactions and dissipation~\cite{Ritsch:2013,Mivehvar:2021,Defenu:2023,Defenu:2024} which enables the atoms to form new phases of matter~\cite{Baumann2010-uv,Dimer:2007,Habibian:2013,Landig:2016,Dogra:2016,Halati:2020,Sharma:2022,Helson:2023,Mivehvar:2019,Marsh:2024,Guo:2021,kroeze:2023,Zhang:2021,Schuster:2020,chelpanova:2024,Larson:2008,Jaeger:2019,Jaeger:2020} and exhibit dynamical features including fast relaxation~\cite{Plestid:2018,Jaeger:2016,Keller:2018,zwettler:2024,Marijanovic:2024}, prethermal dynamics~\cite{Schuetz:2014,Schuetz:2015,Schuetz:2016,Gupta:2016,Wu:2023}, and long-lived oscillations~\cite{Halati:2024,Iemini:2018,Carollo:2022}. The plethora of effects found in such systems along with their simplicity makes them an ideal platform to explore and describe new physics.

Several works~\cite{Piazza:2015,Kessler:2020,Kessler:2021,Dreon:2022,Rosa-Medina:2022,Lin:2022,Skulte:2024,Kongkhambut:2022,Jaeger:2023,Gong:2018,Moodie:2018,Stitely:2020,Stitely:2020:1,Adiv:2024,Dogra:2019} in this field study the emergence of dynamical phases where a true stationary state is not reached on experimentally relevant timescales. These phases have been connected to time-crystals~\cite{Iemini:2018,Gong:2018,Kessler:2020,Kongkhambut:2022,kongkhambut:2024} and require a combination of long-range interactions, non-linear feedback \textit{e.g.} due to a dynamical Stark shift which depends on the atomic pattern, and dissipation from photon losses. Such phases have been reported for blue~\cite{Piazza:2015,Kongkhambut:2022} and red atom-laser detuning~\cite{Gao:2023,Skulte:2024}. However, the mechanisms that result in the emergence and stabilization of such dynamical phases are still not fully understood. In particular, it is an open question whether such dynamical phases can exist in the presence of a large timescale separation between the atomic and cavity degrees of freedom, where the latter can be adiabatically eliminated from the dynamics.

In this paper we explore the underlying mechanisms that results in the emergence and stabilization of dynamical phases. What is key for the existence of these dynamical phases is the optomechanical resonance that emerges because the atoms act as a dynamical refractive index that shifts the cavity frequency, such that certain patterns can compensate for the bare pump-cavity detuning.
We assume the so-called bad cavity regime where the cavity relaxation rate is the largest characteristic frequency of the dynamics.
We derive a mean-field Hamiltonian for the atomic degrees of freedom using perturbation theory in the atom-cavity coupling. In second-order this model includes the effects of the optomechanical frequency shift and of the cavity dissipation. We determine the steady state solution and the dynamical instability in this regime, which we denote by ``adiabatic''. We then include the next-order corrections (non-adiabatic corrections) and identify regimes where the adiabatic approximation fails.
In particular, we find limit cycle phases that undergo a stable dynamical orbit only in the instability region that exists when including non-adiabatic corrections. Surprisingly, the adiabatic mean-field Hamiltonian is sufficient to describe the dynamics of this phase indicating that the limit cycle phase is a metastable configuration that coexists with the ordinary stationary self-organized state. The role of non-adiabatic  
effects is to destabilize the stationary state and to damp excitations around the limit cycle solutions. 

Our paper is structured as follows. In Sec.~\ref{Sec:2} we introduce the theoretical description of the coupled atom-cavity system. In addition, we establish the parameter regime and present the phase and stability diagrams. Next, in Sec.~\ref{Sec:3} we show the real time quench dynamics of the rescaled intensity for the full, adiabatic, and non-adiabatic descriptions. We also demonstrate the metastability of self-organized and limit cycle states when considering adiabatic dynamics. Finally, in Sec.~\ref{Sec:4} we summarize our results and provide a brief outlook.
\section{Theoretical Description\label{Sec:2}}
    \subsection{Model}
    \begin{figure}
\centerline{\includegraphics[width=0.6\linewidth]{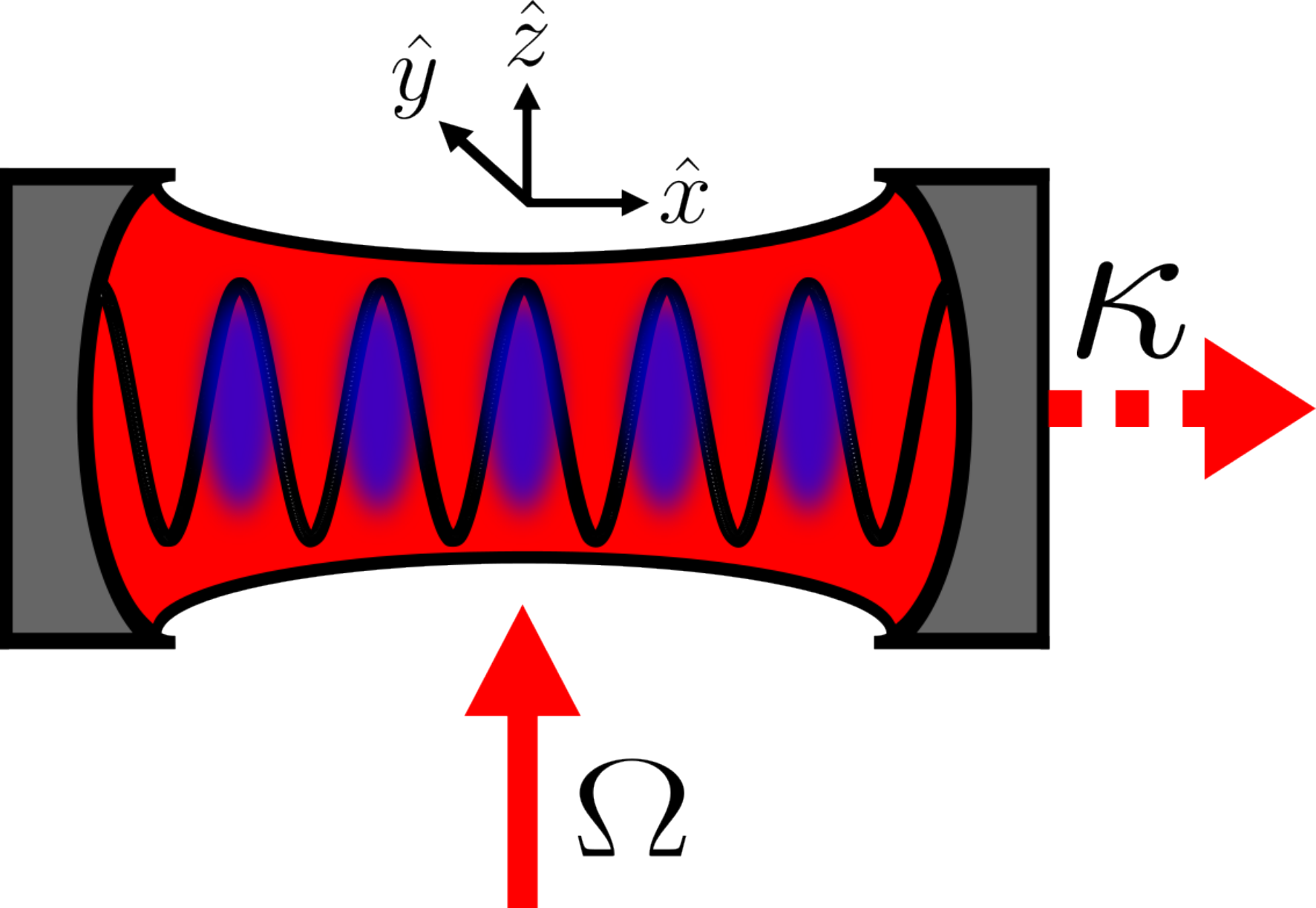}}
        \caption{ Schematic of a experimentally realizable setup. A BEC of $^{87}\text{Rb}$ in the dispersive regime is trapped inside an optical cavity and is transversely driven by a coherent pump laser with Rabi frequency $\Omega$ that is far red-detuned with respect to the atomic transition of interest. The BEC is coupled to a single dissipative mode with decay rate $\kappa$.}
        \label{fig:1}
    \end{figure}
    We consider a BEC that is transversely driven by coherent light far red-detuned with respect to an atomic transition (Fig.~\ref{fig:1}). In this work we study the dynamics of the BEC along the cavity $x$-axis which is strictly justified if the atoms are tightly trapped along the two spatial dimensions transverse to the cavity $x$-axis. The atoms interact with a single-mode of a high-finesse cavity with a standing-wave cavity mode function proportional to the atom-cavity coupling strength $\mathcal{G}_0\text{cos}(k_cx)$, where $\mathcal{G}_0$ is the vacuum Rabi frequency and $k_c$ is the cavity wave number. We work in the dispersive regime which requires a pump detuning $\abs{\Delta_a} = \abs{\omega_p - \omega_a} \gg \gamma$, where $2\gamma$  is
    the full atomic linewidth at half maximum.  In this regime the atomic excited state will be negligibly populated and can be eliminated resulting in the coherent scattering of pump photons into the cavity. The Hamiltonian in second quantization is given by~\cite{Ritsch:2013}
    \begin{equation}
        \begin{aligned}\label{eq:ManybodyHam}
            \hat{\mathcal{H}} = -\hbar\Delta_c\hat{a}^\dagger\hat{a} + \hat{\mathcal{H}}_{\mathrm{kin}} + \hbar U_0\hat{a}^\dagger\hat{a}\hat{\mathcal B} + \hbar\eta(\hat{a}^\dagger + \hat{a})\hat{\vartheta},
        \end{aligned}
    \end{equation}
    and is reported in a reference frame rotating at the pump laser frequency $\omega_p$.
    Here, $\Delta_c = \omega_p - \omega_c$ is the bare cavity detuning with respect to the pump, $\hat{a}^\dagger (\hat{a})$ is the bosonic creation (annihilation) operator for the cavity photons, $U_0 = \mathcal{G}_0^2/\Delta_a$  is the dynamical Stark shift and $\eta = \Omega \mathcal{G}_0/\Delta_a$ is the coherent scattering rate of pump photons into the cavity. We further introduced the kinetic energy
    \begin{align}\label{eq:kineticenergy}
        &\hat{\mathcal H}_{\mathrm{kin}} = \int\hat{\psi}^\dagger(x)\Big(-\frac{\hbar ^2}{2m}\frac{\partial^2 }{\partial x^2}\Big)\hat{\psi}(x)dx,
    \end{align}
    where $m$ is the atomic mass and $\hat{\psi}^\dagger(x) \,(\hat{\psi}(x))$ is the atomic field operator that creates (annihilates) a particle at position $x$. The central quantities for describing the collective effects of the atomic densities are the bunching parameter
    \begin{align}
        &\hat{\mathcal B} = \int\hat{\psi}^\dagger(x)\text{cos}^2(k_cx)\hat{\psi}(x)dx,\label{eq:mathcalB}
    \end{align}
    and order parameter
    \begin{align}
        &\hat{\vartheta} = \int\hat{\psi}^\dagger(x)\text{cos}(k_cx)\hat{\psi}(x)dx.\label{eq:vartheta}
    \end{align}
    A large value of $\langle\hat{\mathcal{B}}\rangle_{\text{MB}}>1/2$ implies that the bunching of atoms are close to the field antinodes. In addition, when the order parameter is  $|\langle\hat{\vartheta}\rangle_{\text{MB}}|>0$ the atoms will form an effective Bragg grating ($\lambda_c$-pattern) in the cavity which supports constructive interference of scattered pump photons. Here, $\langle .\rangle_{\text{MB}}$ denotes the expectation value taken over the many-body state of the atoms and cavity. We mention at this point that both of the atom-cavity couplings proportional to $U_0$ and $\eta$ induce effective atom-atom interactions mediated by the cavity photons. Additional atom-atom interactions such as s-wave collisions are not taken into account in our paper. This approximation allows us to focus on the dynamics that emerge from the cavity-mediated forces.
    \subsection{Mean-field theory}
    We adopt a mean-field description to model the dynamics of the coupled atom-cavity system with a generalized Schrödinger equation
    \begin{equation} \label{eq3}
        \begin{aligned}
            i\hbar\frac{\partial\ket{\psi}}{\partial t} = \hat{H}_{\mathrm{mf}}(\alpha)\ket{\psi},
        \end{aligned}
    \end{equation}
    where we introduce the mean-field Hamiltonian
    \begin{equation}
        \begin{aligned}\label{eq:mfh}
            \hat{H}_{\mathrm{mf}}(\alpha)=\frac{\hat{p}^2}{2m}+\hbar NU_0|\alpha|^2\hat{B}+\hbar\sqrt{N}\eta(\alpha+\alpha^*)\hat{\Theta},
        \end{aligned}
    \end{equation}
    and $\hat{p}=-i\hbar\partial/\partial x$ is the momentum operator.
    In this equation we have introduced the mean-field wave function $\bra{x}\ket{\psi}=\langle\hat{\psi}(x)\rangle_{\mathrm{MB}}/\sqrt{N}$ and rescaled cavity field amplitude $\alpha=\langle\hat{a}\rangle_{\mathrm{MB}}/\sqrt{N}$. We define the single-particle operator describing the bunching parameter $\hat{B}=\cos^2(k_c\hat{x})$ and order parameter $\hat{\Theta}=\cos(k_c\hat{x})$. The mean-field atomic state is coupled dynamically to the cavity field amplitude which evolves according to
    \begin{equation}\label{eq:mfcavity}
        \begin{aligned}
            \frac{\partial \alpha(t)}{\partial t} = (i\delta_c - \kappa)\alpha(t) -i\sqrt{N}\eta\Theta.
        \end{aligned}
    \end{equation}
    Here, we introduced the effective cavity detuning
    \begin{align}
        \delta_c=\Delta_c - NU_0B,\label{eq:deltac}
    \end{align}
    and expectation values $B=\langle \hat{B}\rangle$, $\Theta=\langle\hat{\Theta}\rangle$, where $\langle \hat{O}\rangle=\bra{\psi}\hat{O}\ket{\psi}$ for any single-particle operator $\hat{O}$.  Within the mean-field treatment we assume that we can factorize second-order moments of annihilation and creation operators as products of first-order moments. Also, the expectation values of $B=\langle\hat{\mathcal{B}}\rangle_{\text{MB}}/N$, $\Theta=\langle\hat{\vartheta}\rangle_{\text{MB}}/N$, and $\bra{x}\ket{\psi}=\langle\hat{\psi}(x)\rangle_{\mathrm{MB}}/\sqrt{N}$ are a consequence of the definitions in Eqs.~\eqref{eq:mathcalB} and \eqref{eq:vartheta}.
    \subsection{Parameter regime}
    In this paper we want to analyze the onset of dynamical phases for this atom-cavity system. To see their existence we require that the atom-cavity system is close to optomechanical resonance which is determined by $\delta_c=0$. Assuming the resonant case we can calculate the corresponding value of $B=\Delta_c/(NU_0)$. Note that $B$ being $0<B<1$ shows that it is important that $|\Delta_c|\leq|NU_0|$. In contrast to Ref.~\cite{Piazza:2015}, but similar to Ref.~\cite{Gao:2023} we work in the regime of red atom-pump detuning $\Delta_a<0$ and red bare cavity detuning $\Delta_c<0$. As a consequence, the dynamical Stark shift is negative, $U_0<0$. This means that in the presence of a nonzero cavity field the atoms can minimize their energy by localizing close to the antinodes of the cavity mode profile. To garner an intuition of this process it is convenient to set $\partial\alpha/\partial t=0$ to solve for the stationary state
    \begin{equation}
        \begin{aligned}\label{eq5}
            \alpha_0 &= \frac{\sqrt{N}\eta \Theta}{\delta_c + i\kappa}.
        \end{aligned}
    \end{equation}
    Substituting this result in the Hamiltonian~\eqref{eq:mfh}
    we find
    \begin{equation}
        \begin{aligned}\label{eq:mfhelcav}
\hat{H}_{\mathrm{mf}}(\alpha_0)=\frac{\hat{p}^2}{2m}+\frac{\hbar NU_0N\eta^2\Theta^2}{\delta_c^2+\kappa^2}\hat{B}+\frac{2\hbar \delta_c N\eta^2\Theta}{\delta_c^2+\kappa^2}\hat{\Theta}.
        \end{aligned}
    \end{equation}
    For our parameters we have in general $\delta_c<0$ and $U_0<0$ such that the mean-field potential energy can be reduced by increasing $\Theta^2$ and $\hat{B}$. Thus a $\delta_c<0$ favors organized and bunched atomic ensembles. However, an increasing value of $B$ can result in a blue-shifted value of $\delta_c$ [Eq.~\eqref{eq:deltac}] making the organization of the atomic ensemble less ``attractive''. This negative feedback loop has been shown to give rise to dynamical and oscillatory phases that do not reach a true stationary state~\cite{Gao:2023}. In contrast to previous theoretical papers, we study the bad cavity situation where the cavity evolves much faster than the atomic degrees of freedom $\kappa,|\Delta_c|\gg\omega_r,\sqrt{N}\eta$, with recoil frequency $\omega_r=\hbar k_c^2/(2m)$. Working in this regime has two main implications. First, because of sufficiently weak coupling between the cavity and atoms, the occupation of hybrid atom-cavity states is largely suppressed. This implies that one can treat the atoms and cavity to good approximation separately. Second, the relaxation of the cavity is fast and therefore one can eliminate the cavity degrees of freedom from the atomic equations using perturbation theory.  Working in this regime also makes our work experimentally relevant. To demonstrate this we have chosen the parameters to be similar as in the experiment of Ref.~\cite{Baumann2010-uv} for a BEC of $^{87}\text{Rb}$. The parameters we have chosen here correspond to the top left quadrant of Fig. 5(a) in Ref.~\cite{Baumann2010-uv} where the system is close to optomechanical resonance and a frustrated system has been predicted Ref.~\cite{Ritsch:2006}.

      \subsection{Imaginary time phase diagram}
    We follow the general intuition that the underlying mean-field wave function is the ground state of the mean-field Hamiltonian. To find this wave function we employ an imaginary time propagation method (ITPM) along with the adiabatic elimination of the cavity field dynamics. This method has been used in several works~\cite{Gao:2023,Nagy2008,Mivehvar:2019}. The stationary field of the photons is given by Eq.~\eqref{eq5} and by substituting $\tau = it$, Eq.~\eqref{eq3} becomes
    \begin{equation}
        \begin{aligned}\label{eq6}
            \hbar\frac{\partial \ket{\psi}}{\partial \tau} = -\Delta\hat{H}_{\mathrm{mf}}\ket{\psi}
        \end{aligned}
    \end{equation}
    with $\Delta\hat{H}_{\mathrm{mf}}=\hat{H}_{\mathrm{mf}}(\alpha_0) - \expval{\hat{H}_{\mathrm{mf}}(\alpha_0)}$ and the mean-field Hamiltonian given by Eq.~\eqref{eq:mfhelcav}. Evolving Eq.~\eqref{eq6} over long imaginary times $\tau$ results in what we expect to be the ground state. We comment at this point that Eq.~\eqref{eq6} guides the dynamics to a stationary state of the mean-field equation which is an eigenstate of $\hat{H}_{\mathrm{mf}}(\alpha_0)$.  The properties of the numerically found eigenstates are discussed in the following.

    \begin{figure*}[ht]
        \centering
        \includegraphics[width=1.0\textwidth]{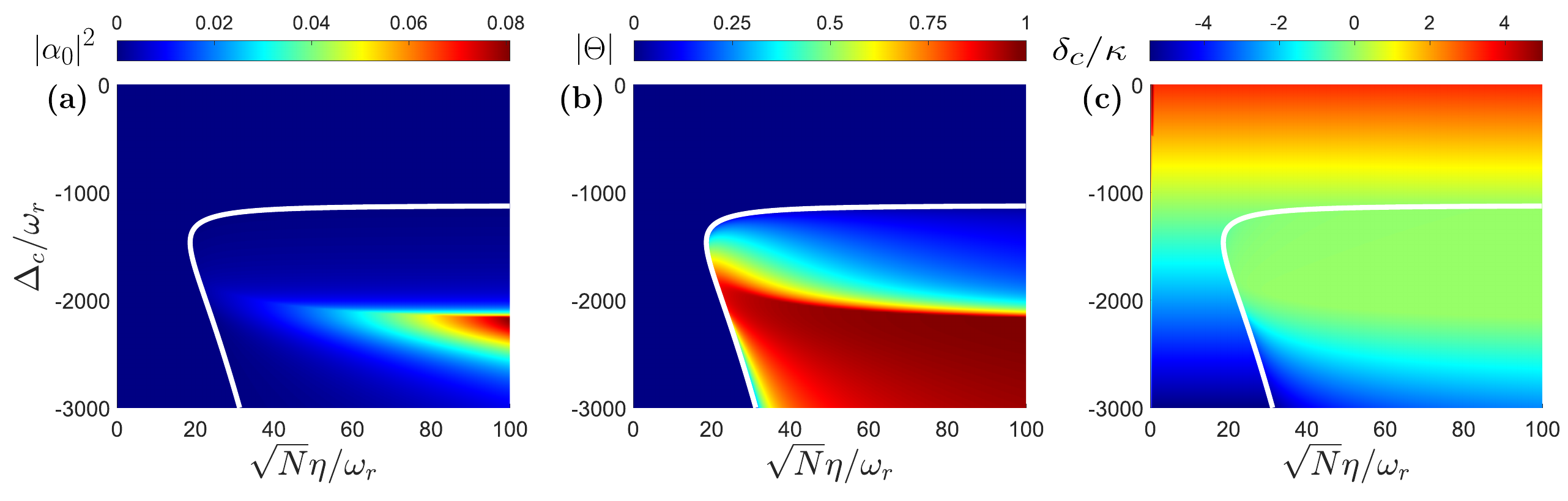}
        \caption{Steady state imaginary time phase diagrams of the (a) rescaled cavity field intensity $\abs{\alpha_0}^2$, (b) order parameter $\abs{\Theta}$, and (c) effective cavity detuning $\delta_c/\kappa$ as a function of the bare cavity detuning $\Delta_c$ and effective cavity pump strength $\sqrt{N}\eta$. We vary $\sqrt{N}\eta$ and fix $N$ keeping $NU_0 = -2241.38\omega_r$ constant and $\kappa = 344.83\omega_r$. The solid white line denotes the phase boundary from the unorganized to self-organized phase [Eq.~\eqref{last}]. We choose a momentum cutoff of 20$\hbar k_c$.}
        \label{fig3}
    \end{figure*}
    We now present the steady state mean-field imaginary time phase diagrams of the rescaled cavity field intensity $|\alpha_0|^2$, order parameter $|\Theta|$, and effective cavity detuning $\delta_c$ as a function of the bare cavity detuning and effective cavity pump strength $\sqrt{N}\eta$. In this paper we will vary $\sqrt{N}\eta$ keeping the atom number $N$ constant. In an experiment that corresponds to changing the driving laser intensity. Above a critical pump strength defined by 
    \begin{equation}
        \begin{aligned}\label{last}
            \sqrt{N}\eta_c = \sqrt{\frac{(\Delta_c - NU_0/2)^2 + \kappa^2}{{(NU_0 - 2\Delta_c)}}}\sqrt{\omega_r},
        \end{aligned}
    \end{equation}
    which has been derived in Refs.~\cite{Gao:2023,Nagy2008}, we expect the system to be self-organized. For small values of $\sqrt{N}\eta$ and $\Delta_c$ we find a vanishing photon number $|\alpha_0|^2$ [see Fig.~\ref{fig3}(a)] and order parameter $\Theta$ [see Fig.~\ref{fig3}(b)]. Beyond the threshold marked by Eq.~\eqref{last} we find a non-vanishing photon number and self-organized density grating $\Theta\neq0$. In Fig.~\ref{fig3}(c) we see how the effective cavity detuning $\delta_c = (\Delta_c - NU_0{B})$ changes in the $\sqrt{N}\eta-\Delta_c$ plane. Outside of the threshold the system remains in a homogeneous BEC with a fixed bunching parameter of ${B} = 1/2$. Which results in $\delta_c$ being invariant under changes in $\sqrt{N}\eta$. Within the threshold we see for a large range of $\Delta_c$ and $\sqrt{N}\eta$, the effective cavity detuning $\delta_c$ is near zero. This means the bare cavity detuning and the dynamical Stark shift are of equal value, i.e., $\Delta_c \approx NU_0{B}$. Interestingly, the atoms organize into patterns which satisfy the optomechanical resonance and therefore the state found by the imaginary time evolution moves onto optomechanical resonance until $\Delta_c \approx NU_0$.
    Beyond this line, for $\Delta_c>NU_0$, the system becomes strongly organized as seen in Fig.~\ref{fig3}(b). The formation of a deep Bragg grating, $|\Theta|\lessapprox1$, requires the dispersive cavity-mediated forces to be mostly pronounced which appears in the regime $-\delta_c\gg\kappa$ [Fig.~\ref{fig3}(c)]. The cavity field intensity [Fig~\ref{fig3}(a)] exhibits a maximum around $\Delta_c\sim NU_0$ due to a combination of a highly organized atomic ensemble $|\Theta| \approx 1$ and a large value of $1/(\delta_c^2+\kappa^2)$ [Eq.~\eqref{eq5} for comparison].

    In this paper we are interested in the dissipative regime i.e., $\delta_c \ll \kappa$ where the effective cavity detuning vanishes due to the dynamical Stark shift, causing a negative feedback between $\Theta$ and ${B}$. In this region dynamical phases have been found and in order to probe dynamical instabilities within the organized phase, we utilize the steady state solution found by the ITPM and perform a stability analysis with it. This analysis is presented in the following subsection.
    \subsection{Stability analysis}
    \begin{figure}[ht]
        \centering
        \includegraphics[width=1.0\columnwidth]{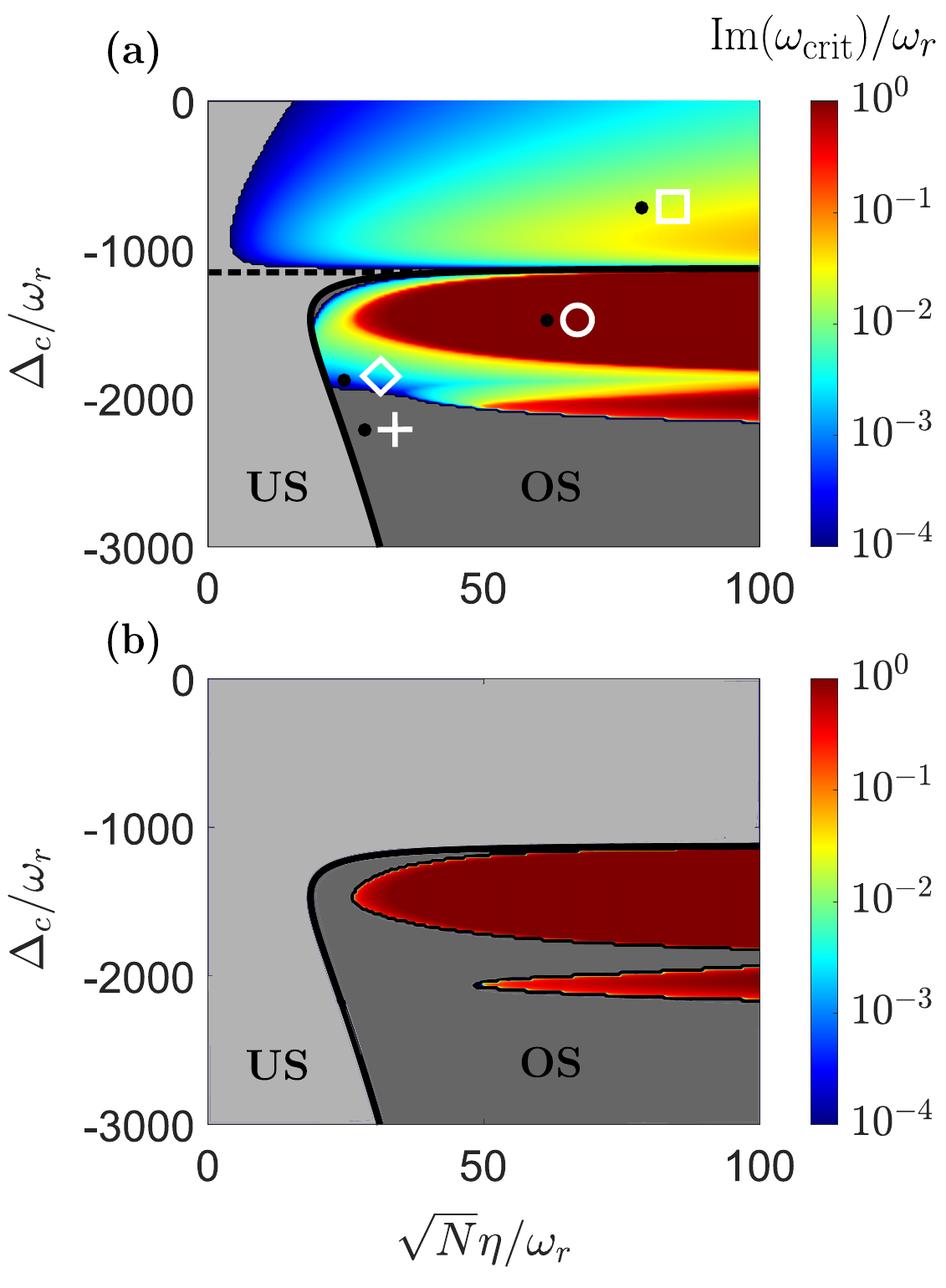}
        \caption{(a) Maximum imaginary eigenvalue Im$(\omega_{\mathrm{crit}})$ determined from Eq.~\eqref{eq19} as a function of the effective cavity pump strength $\sqrt{N}\eta$ and bare cavity detuning $\Delta_c$. (b) Maximum imaginary eigenvalue determined from Eq.~\eqref{eq21}. The light grey shaded regions represent the unorganized stable phase (US) and the darker grey shaded regions represent the organized stable phase (OS). The black dashed line is at $\Delta_c = NU_0/2$ and the black contour is determined by Eq.~\eqref{last}. We vary $\sqrt{N}\eta$ and keep $N$ constant and use $NU_0 = -2241.38\omega_r$ and $\kappa = 344.83\omega_r$. We choose a momentum cutoff of 20$\hbar k_c$.}
        \label{fig4}
    \end{figure}
    In this section, we study the stability of a steady state of the ITPM given by Eq.~\eqref{eq6}. It is important to note that the  imaginary time solutions will also be stationary with respect to the real time evolution described by Eqs.~\eqref{eq3}-\eqref{eq5}. However, they are not necessarily stable against fluctuations. The stability gives us insight about dynamical solutions within the organized phase when studying their real time dynamics. This is done by linearizing Eqs.~\eqref{eq3}-\eqref{eq5} for small fluctuations around the large steady state solution $\alpha_0$ and $\ket{\psi_0}$. These fluctuations are denoted by $\delta\alpha = \alpha - \alpha_0$ and $\ket{\delta\psi} = e^{i\mu t}\ket{\psi} - \ket{\psi_0}$, where $\mu=\expval{\hat{H}_{\mathrm{mf}}(\alpha_0)}$ is the chemical potential. This approach is valid if fluctuations in the cavity field and atomic degrees of freedom are small.
    The linearized equations of motion of the fluctuations in $\ket{\delta\psi}$ and $\delta\alpha$ are given by
    \begin{equation}\label{eq15}
        \begin{aligned}
            i\hbar\frac{\partial \ket{\delta\psi}}{\partial t} =& \Delta\hat{H}_{\mathrm{mf}}\ket{\delta\psi}+\hbar(\hat{S}^\dag\delta\alpha+\delta\alpha^*\hat{S})\ket{\psi_0},
        \end{aligned}
    \end{equation}
where $\hat{S}=\sqrt{N}\eta\hat{\Theta}+NU_0\alpha_0\hat{B}$ and
    \begin{equation}\label{eq16}
        \begin{aligned}
            i\frac{\partial \delta \alpha}{\partial t} &= \left(-\delta_c - i\kappa\right)\delta\alpha +\left(\bra{\delta\psi}\hat{S}\ket{\psi_0}+\bra{\psi_0}\hat{S}\ket{\delta\psi}\right), 
        \end{aligned}
    \end{equation}
    with $\delta_c=\Delta_c-NU_0\bra{\psi_0}\hat{B}\ket{\psi_0}$.
    Since the linearized equations couples $\ket{\delta\psi}$ and $\delta\alpha$ to their complex conjugates, we make the ansatz
    \begin{equation}\label{eq18}
        \begin{aligned}
            \ket{\delta\psi} =& e^{-i\omega t}\ket{\delta\psi_+} + e^{i\omega^* t}\ket{\delta\psi_-^*},\\ \delta\alpha =& e^{-i\omega t}\delta\alpha_+ +  e^{i\omega^* t}\delta\alpha^*_-.
        \end{aligned}
    \end{equation}
    Here we introduced the complex conjugate wave function which is defined as $\bra{x}\ket{\psi^*}=\bra{\psi}\ket{x}$ and $()^*$ denotes the complex conjugation.
    Using this ansatz we can derive a linear eigenvalue problem
    \begin{equation}\label{eq19}
        \begin{aligned}
            \omega \begin{pmatrix} \delta\alpha_+ \\ \delta\alpha_- \\ \ket{\delta\psi_{+}} \\ \ket{\delta\psi_{-}} \end{pmatrix} = \hat{\textbf{M}}_1\begin{pmatrix} \delta\alpha_+ \\ \delta\alpha_- \\ \ket{\delta\psi_{+}} \\ \ket{\delta\psi_{-}} \end{pmatrix},
        \end{aligned}
    \end{equation}
    where $\hat{\textbf{M}}_1$ is a non-Hermitian matrix determined from Eqs.~\eqref{eq15} and \eqref{eq16}.  The matrix is found using Eq.~\eqref{eq18} in Eqs.~\eqref{eq15} and \eqref{eq16} and given by
    \begin{align}
        \hat{\bf M}_1=\begin{pmatrix}
            -\delta_c - i\kappa &0&\bra{\psi_0}\hat{S}&\bra{\psi_0^*}\hat{S}\\
0&\delta_c-i\kappa&-\bra{\psi_0}\hat{S}^\dag&-\bra{\psi_0^*}\hat{S}^\dag\\
\hat{S}^\dag\ket{\psi_0}&\hat{S}\ket{\psi_0}&\Delta\hat{H}_{\mathrm{mf}}/\hbar&0\\
-\hat{S}^\dag\ket{\psi_0^*}&-\hat{S}\ket{\psi_0^*}&0&-\Delta\hat{H}_{\mathrm{mf}}/\hbar&
        \end{pmatrix}.
    \end{align}
Note that we have explicitly used that the complex conjugate operation acts trivially on the Hamiltonian $\hat{H}_{\mathrm{mf}}$ and the operators $\hat{B}$ and $\hat{\Theta}$ such that $\hat{S}^*=\hat{S}^\dag$.

The eigenvalues of $\hat{\bf M}_1$ describe the dynamics of fluctuations around the mean-field steady state $\ket{\psi_0}$ and $\alpha_0$. In particular, this enables us to access the fluctuation dynamics of the atomic and cavity degrees of freedom on equal timescales. 

Since our work is mostly focused on the regime where we have a clear timescale separation, we derive in the following the stability assuming that fluctuations of the cavity field instantaneously follow the fluctuations of the atomic degrees. A comparison of both approaches, with and without the adiabatic assumption, can therefore shed light on which effects emerge from either the non-adiabatic corrections or from the pure adiabatic assumption. Envoking the adiabatic assumption we can use
    \begin{equation}
        \begin{aligned}\label{eq20}
            \delta\alpha &= \frac{\bra{\delta\psi}\hat{S}\ket{\psi_0}+\bra{\psi_0}\hat{S}\ket{\delta\psi}}{\delta_c + i\kappa} .
        \end{aligned}
    \end{equation}
    Now substituting Eq.~\eqref{eq20} into Eq.~\eqref{eq15} one gets a reduced eigenvalue problem
    \begin{equation}
        \begin{aligned}\label{eq21}
            \omega \begin{pmatrix} \ket{\delta\psi_+} \\ \ket{\delta\psi_-}\end{pmatrix} = \hat{\textbf{M}}_2 \begin{pmatrix}
                \ket{\delta\psi_+} \\ \ket{\delta\psi_-}
            \end{pmatrix},
        \end{aligned}
    \end{equation}
    with 
    	\begin{align}
    	\hat{\bf M}_2=\begin{pmatrix}
    	\Delta\hat{H}_{\mathrm{mf}}/\hbar+\hat{A}&\hat{B}\\
    	-\hat{B}^*&-[\Delta\hat{H}_{\mathrm{mf}}/\hbar+\hat{A}^*]&
    	\end{pmatrix},
    	\end{align}
        	\begin{align}
    \hat{A}=&\frac{\hat{S}^\dag\ket{\psi_0}\bra{\psi_0}\hat{S}}{\delta_c+i\kappa}\nonumber+\frac{\hat{S}\ket{\psi_0}\bra{\psi_0}\hat{S}^\dag}{\delta_c-i\kappa},\end{align}
    \begin{align}
    \hat{B}=&\frac{\hat{S}^\dag\ket{\psi_0}\bra{\psi_0^*}\hat{S}}{\delta_c+i\kappa}+\frac{\hat{S}\ket{\psi_0}\bra{\psi_0^*}\hat{S}^\dag}{\delta_c-i\kappa}.
    \end{align}
    
    In what follows we will numerically study the stability of fluctuations based on the theory where the cavity is included, i.e., the eigenvalues of $\hat{\bf M}_1$ [Eq.~\eqref{eq19}]. We then compare the results with the stability determined by the eigenvalues of $\hat{\bf M}_2$ describing the adiabatically eliminated cavity degrees of freedom [Eq.~\eqref{eq21}].
    
    The solution $\ket{\psi_0}$ and $\alpha_0$ is stable if $\mathrm{Im}(\omega)<0$ for all eigenvalues of $\omega$. For this we first find $\ket{\psi_0}$ and $\alpha_0$ using the ITPM and then find the eigenvalues of $\hat{\bf M}_i$ ($i=1,2$) numerically. We then determine the eigenvalue $\omega_\mathrm{crit}$ with the largest imaginary part, which corresponds to the 
    ``most unstable'' or ``least stable'' mode. Because of the numerical limitations in finding the steady state  and eigenvalues due to the discretization of momentum states and time steps, we call the system unstable if $\mathrm{Im}(\omega_{\mathrm{crit}})>10^{-4}\omega_r$. The regions of instability are visible in color in Fig.~\ref{fig4}(a) and (b) for the non-adiabatic case ($\hat{\bf M}_1$) and adiabatic case ($\hat{\bf M}_2$), respectively. The regions where $\mathrm{Im}(\omega_{\mathrm{crit}})<10^{-4}\omega_r$ is visible in grey colors.
    
     First, below threshold [outside the black contour determined by Eq.~\eqref{last}], the steady state is given by $\alpha_0 = 0$ and $\ket{\psi_0}=\ket{p=0}$ (the zero-momentum state). Already below threshold we find a region of instability in Fig.~\ref{fig4}(a) which is not visible in Fig.~\ref{fig4}(b). The zero-momentum state and $\alpha_0=0$ is unstable for $\delta_c=\Delta_c-NU_0/2>0$ or equivalently $\Delta_c>NU_0/2$. The reason for this is a blue effective cavity detuning which results in heating of the atomic ensemble. This heating effect is not described by the adiabatic elimination and the reason why this instability region is seen can be attributed to the non-adiabatic terms simulated in Fig.~\ref{fig4}(a) but not in Fig.~\ref{fig4}(b). We remark at this point that the grey area where $\mathrm{Im}(\omega_{\mathrm{crit}})<10^{-4}$ but $\Delta_c>NU_0/2$ is a numerical artifact since the heating rate becomes very small for small values of $\sqrt{N}\eta$. In fact, we expect heating for all values of $\sqrt{N}\eta>0$ provided $\Delta_c>NU_0/2$. For regions where $\Delta_c<NU_0/2$ and $\sqrt{N}\eta$ below threshold we denote the phase as unorganized stable (US).
     
     When crossing the transition line we find a large parameter space in Fig.~\ref{fig4}(b) where the superradiant solution $\alpha_0\neq0$ is stable (visible as dark grey). This phase is called organized stable (OS). In addition, we find two regions of instability which appear for parameters $NU_0/2>\Delta_c>NU_0$. In this region the steady state also exhibites a very small effective cavity detuning $\delta_c\approx0$ [Fig.~\ref{fig3}(c)]. Remarkably, the instability region is significantly extended when including non-adiabatic corrections as visible in Fig.~\ref{fig4}(a). In that case we find an instability almost everywhere across the phase transition provided that $NU_0/2>\Delta_c>NU_0$, in the regime where we have found optomechanical resonance with the imaginary-time evolution. We remark that this finding is in qualitative agreement with Ref.~\cite{Gao:2023} and one can expect enhanced cavity field fluctuations since this is the typical region of bistability.
     The fact that we find small detuning $\delta_c$ [Fig.~\ref{fig3}(c)] in this region also implies a pronounced role of dissipation in the form of photon losses. Thus the full dissipative response to the atoms is only correctly described when including beyond adiabatic effects. This can be attributed to enhanced fluctuations in the cavity field that are not captured by the adiabatic elimination within the region of bistability. In this context the results visible in Fig.~\ref{fig4}(b) show how far one can push the adiabatic elimination for studying the linear stability close to the edge of a bistable region.
     It is important to note, however, the large difference in magnitude of $\mathrm{Im}(\omega_\mathrm{crit})$ when comparing the instability regions of Fig.~\ref{fig4}(a) and (b). Within the instability regions due to the non-adiabatic corrections in (a) but not in (b) we see the build up of fluctuations takes several orders of magnitude longer than the red instability regions that are present in both (a) and (b). For an actual experiment this could mean that although the region is unstable the build up of fluctuations might not be observed on experimentally relevant timescales.  
     
     Our analysis has demonstrated that the mean-field steady states are unstable in certain parameter regions. Although we know that the initial dynamics would be governed by an exponential increase, our theory can so far not predict what happens on longer timescales. For this we provide an analysis of the mean-field real time dynamics in the following section.
    \section{Dynamics \label{Sec:3}}
    	In this section we study the real time dynamics of the coupled atom-cavity system. For this our main tools are the numerical integration of Eq.~\eqref{eq3} and different approximations of the cavity field in Eq.~\eqref{eq:mfcavity}.
	
	\subsection{Quench dynamics}
	\begin{figure}
		\centering
		\includegraphics[width=\columnwidth]{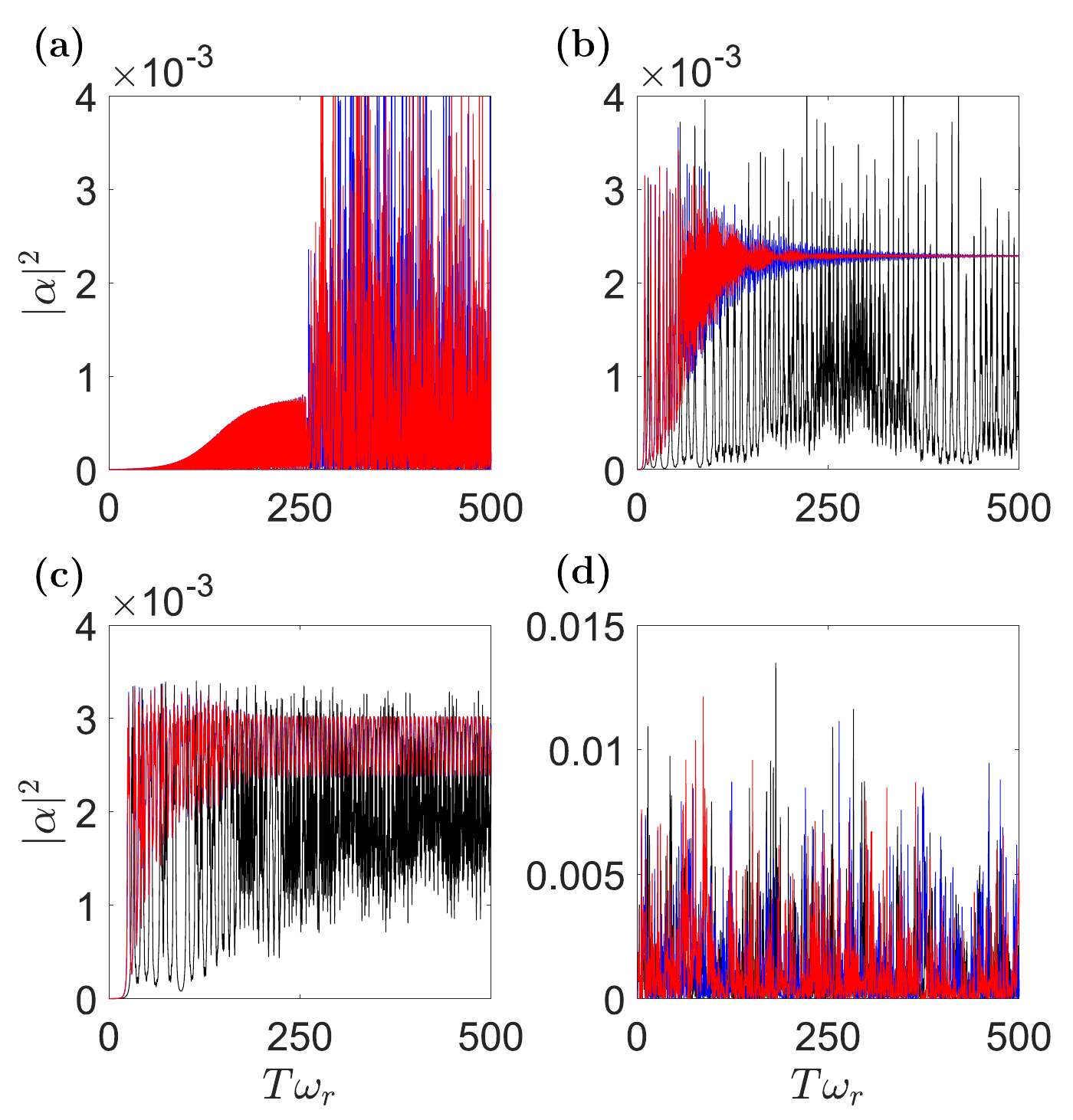}
		\caption{Real time quench dynamics of the rescaled cavity field intensity $\abs{\alpha}^2$. (a) Heating phase with $\Delta_c = -700\omega_r$ and $\sqrt{N}\eta = 80\omega_r$ $[\square-$symbol in Fig.~\ref{fig4}(a)], (b) stationary self-organized phase with $\Delta_c = -2150\omega_r$ and $\sqrt{N}\eta = 27\omega_r$ [$\boldsymbol{+}$-symbol in Fig.~\ref{fig4}(b)], (c) limit cycle phase with $\Delta_c = 1900\omega_r$ and $\sqrt{N}\eta = 22\omega_r$ $[\boldsymbol{\diamond}-$symbol in Fig.~\ref{fig4}(a)], and  (d) chaotic phase with $\Delta_c = 1450\omega_r$ and $\sqrt{N}\eta = 60\omega_r$ $[\boldsymbol{\circ-}$symbol in Fig.~\ref{fig4}(a)]. For the red lines we simulate the cavity field expression of \eqref{eq:mfcavity}, the black lines we simulate Eq.~\eqref{eq5}, and for the blue lines we simulate Eq.~\eqref{eq23}. The other parameters are the same as in Figs.~\ref{fig3}.}
		\label{fig5}
	\end{figure}
	We analyze the real time quench dynamics of an atom-cavity system that is initialized in the zero momentum state, $\ket{\psi(t=0)}=\ket{p=0},$ with an empty cavity field, $\alpha(t=0)=0$. After this initialization we suddenly quench the system parameters to various parameter points of $\sqrt{N}\eta$ and $\Delta_c$ as signified in Fig.~\ref{fig4}(a). In Fig.~\ref{fig5} we show the resulting dynamics of the rescaled cavity field intensity $|\alpha|^2$ as a red solid line for four different points using Eq.~\eqref{eq:mfcavity}. In addition to this red line we have also simulated the dynamics of the cavity field that relies on the adiabatic elimination [Eq.~\eqref{eq5}]. The result of this simulation is shown as a black solid line and will be compared to the full dynamics in the following. The blue curves in Fig.~\ref{fig5} include the non-adiabatic corrections and are discussed at a later point in Sec.~\ref{sec3b}. 
	
	In Fig.~\ref{fig5}(a) we show the dynamics of the cavity field intensity which represents the parameter point where $\sqrt{N}\eta>0$ and $\Delta_c>NU_0/2$ [${\square}$-symbol in Fig.~\ref{fig4}(a)]. Focusing on the red line, we see that within this regime we expect heating to occur due to a blue effective cavity detuning. We observe such an effect in the underlying wave function $\ket{\psi}$ which occupies higher and higher momentum states. In this context we chose an appropriate high cut-off in momentum states for the simulation and timescale shown. The dynamics of the cavity field appears to be rather complex. We first observe a slow increase in the cavity field and after a longer timescale we see an almost chaotic time evolution of the cavity field. The black solid line describing the adiabatic elimination predicts a cavity field that remains at zero for all times. This is consistent with Fig.~\ref{fig4}(b) predicting no instability when studying the adiabatic dynamics for $\Delta_c>NU_0/2$.
	
	Figure~\ref{fig5}(b) represents the parameter point in the OS phase [$\boldsymbol{+}$-symbol in Fig.~\ref{fig4}(a)]. Here, for both the non-adiabatic and adiabatic time evolutions we expect the initial zero momentum state to be unstable. The dynamics of the cavity field are still very different for the red and black lines on long timescales. Both show a very fast initial rise of the cavity field. However, while the red line converges over a long timescale a similar relaxation to a steady state is not visible for the black line. This highlights that the non-adiabatic corrections are important for the relaxation process on longer timescales. After a sudden quench we have injected excitations into the system which can only be damped when including these non-adiabatic corrections. We have also compared the stationary state of the real time evolution with the stationary state of the imaginary time evolution $\ket{\psi_0}$, $\alpha_0$ and found very good agreement. That indicates that the long-time stationary state in this regime after a sudden quench is the one found by the imaginary time. 
	
	In Fig.~\ref{fig5}(c) we show the parameter point which is beyond the critical value of $\sqrt{N}\eta$ [Eq.~\eqref{last}] and within the unstable regime of Fig.~\ref{fig4}(a) [$\boldsymbol{\diamond}$-symbol] but in the stable regime of Fig.~\ref{fig4}(b). For such a parameter point the imaginary time evolution predicts an organized steady state which should be dynamically unstable when studying the full dynamics of the atoms and cavity field. For both simulations, red and black lines, we see an initial quick increase of the photon numbers. This is a consequence of the initial state $\ket{p=0}$ being unstable. On longer timescales for both trajectories we see they do not find a true stationary state. The red curve relaxes towards a stable oscillation which can be understood as a limit cycle phase and the black curve shows relatively chaotic dynamics. While the chaotic dynamics of the black curve is again a consequence of the absence of correct damping in the purely adiabatic theory. The existence of a dynamical phase for the red curve is consistent with the instability of the imaginary time stationary state. These oscillations appear to be very stable, which highlights that certain excitations are imprinted after the sudden quench and can be damped guiding the atom-cavity system into a stable orbit. Such phases have been predicted in this model as seen in Refs.~\cite{Piazza:2015,Gao:2023}, however, here we find them in a regime with a very large timescale separation between the atoms and cavity.
	
	Figure~\ref{fig5}(d) quenches the system into a regime where the stationary state of the imaginary time evolution is unstable for the non-adiabatic and adiabatic theories [$\boldsymbol{\circ}$-symbol in Fig.~\ref{fig4}(a)]. In this regime we find highly chaotic dynamics for the red and black curves. After a rapid initial rise we find irregular and fast oscillations without a clear structure. In such a regime it is questionable if mean-field theory is sufficient to describe these dynamics, because we expect the fluctuations to play an important role. Although we have checked that the momentum cut-off is sufficient we cannot trust this description on long timescales due to heating. 

    We want to emphasize that we work in a regime where the atoms and cavity evolve on very different timescales. In such a regime it should be possible to include non-adiabatic corrections perturbatively into the dynamics. In the next section we derive these corrections and show that they can recapture the main features of the relaxation dynamics as seen with the full cavity description.

	\subsection{Role of non-adiabatic corrections}\label{sec3b}
	We now study how non-adiabatic corrections to the field contribute to the real time quench dynamics. By way of formal integration of Eq.~\eqref{eq:mfcavity} we obtain the expression
	\begin{equation}
		\begin{aligned}\label{eq22}
			\alpha(t) \approx  -i\sqrt{N}\eta\int_0^t e^{{\int_0^{\Tilde{\tau}}}(i[\Delta_c - NU_0B(t - \Tilde{\tau}')] - \kappa)d\Tilde{\tau}'}\Theta(t - \Tilde{\tau})d\Tilde{\tau}.
		\end{aligned}
	\end{equation}
In this equation we have included the explicit time dependence of the bunching parameter and order parameter. In addition we have dropped the initial condition $\alpha(0)$ which requires that the timescale of integration $t$ to be much longer than $1/\kappa$.	We further assume that this timescale $t$ is much shorter than the typical timescale for which $\Theta$ and $B$ evolve, $1/\omega_r\gg t\gg1/\kappa$. Such a coarse graining is possible only in the regime where one has a large separation of timescales, $\kappa\gg\omega_r$.
	Under these assumptions we can Taylor expand the time-dependent terms to first-order and carry out the integration. Arriving at the approximate form of the corrected field
		\begin{equation}
			\begin{aligned}\label{eq23}
				\alpha(t) \approx \frac{i\sqrt{N}\eta\Theta}{i\delta_c-\kappa} + \frac{i\sqrt{N}\eta\dot{\Theta}}{(i\delta_c-\kappa)^2} - \frac{\sqrt{N}\eta NU_0\Theta{\dot{B}}}{(i\delta_c-\kappa)^3}.
			\end{aligned}
		\end{equation}
	For details we refer to App.~\ref{App:A}.	
		
	The expression above contains derivatives of $\Theta$ and $B$. By including these terms in the Hamiltonian $\hat{H}_{\mathrm{mf}}(\alpha(t))$ we incorporate certain retardation effects without explicitly evolving the cavity degrees of freedom. This is a key feature that Eq.~\eqref{eq5} lacks since it does not take into account any retardation effects that arise from the motion of the atoms. We remark that although the time derivatives can be computed numerically one can also extract an analytical formula for them which reads
	\begin{equation}
		\begin{aligned}\label{eq24}
			\dot{\Theta} = \frac{1}{i\hbar}\expval{\big[\hat{\Theta}, \frac{\hat{p}^2}{2m} \big]},\\
			\dot{B} = \frac{1}{i\hbar}\expval{\big[\hat{B}, \frac{\hat{p}^2}{2m} \big]}.
		\end{aligned}
	\end{equation}

 We will now demonstrate that these corrections have a critical influence in recovering the steady state and limit cycle phases. For this we evolve the atomic dynamics using explicitly the ``new'' elimination of the cavity given by Eq.~\eqref{eq23}. The results of the simulations are visible in Fig.~\ref{fig5}(a)-(d) as blue lines.
 
 While the dynamics in Fig.~\ref{fig5}(a) could not be captured by the adiabatic elimination, when including the non-adiabatic corrections we already see very good qualitative and quantitative agreement. The blue line basically captures the slow initial dynamics and the more chaotic long-time dynamics. This also shows that these retardation effects are sufficient to describe the heating and complex dynamics.
 
 In the OS phase [Fig.~\ref{fig5}(b)] we find that the blue curve relaxes to the same stationary state as the red curve. This shows that the first-order retardation effects describe in Eq.~\eqref{eq23} are sufficient for the damping. A closer look shows that the red curve has slightly faster relaxation dynamics. We believe this is due to higher order retardation effects which slightly speed up the damping behavior.
 
 The limit cycle dynamics in Fig.~\ref{fig5}(c) is extremely well reproduced by the blue curve. We only find very small differences between the red and blue curves. This demonstrates that the retardation effects are sufficient to stabilize the limit cycle phase and highlights that these phases can exist despite the large timescale separation.
 
 At last, in Fig.~\ref{fig5}(d) we find very similar dynamics for all three simulation methods via the red, blue, and black curves. We do not find perfect agreement which also means that higher order retardation effects are required for a better quantitative agreement. 
 
 In conclusion, we would like to establish Eq.~\eqref{eq23} as a minimal model to describe the relaxation dynamics in such a system.
 
\subsection{Metastable self-organized and limit cycle states}
Ultimately, we try to obtain a deeper understanding of the parameter regime where the limit cycle phase has been found. From the stability analysis in Fig.~\ref{fig4} we know that in this regime the OS phase is unstable when including the full cavity field, but is stable when only considering purely adiabatic dynamics. The question we seek to ask is whether the adiabatic evolution is able to describe the limit cycle phase at all. 

		\begin{figure}[t]
		\centering
        \includegraphics[width=1.0\columnwidth]{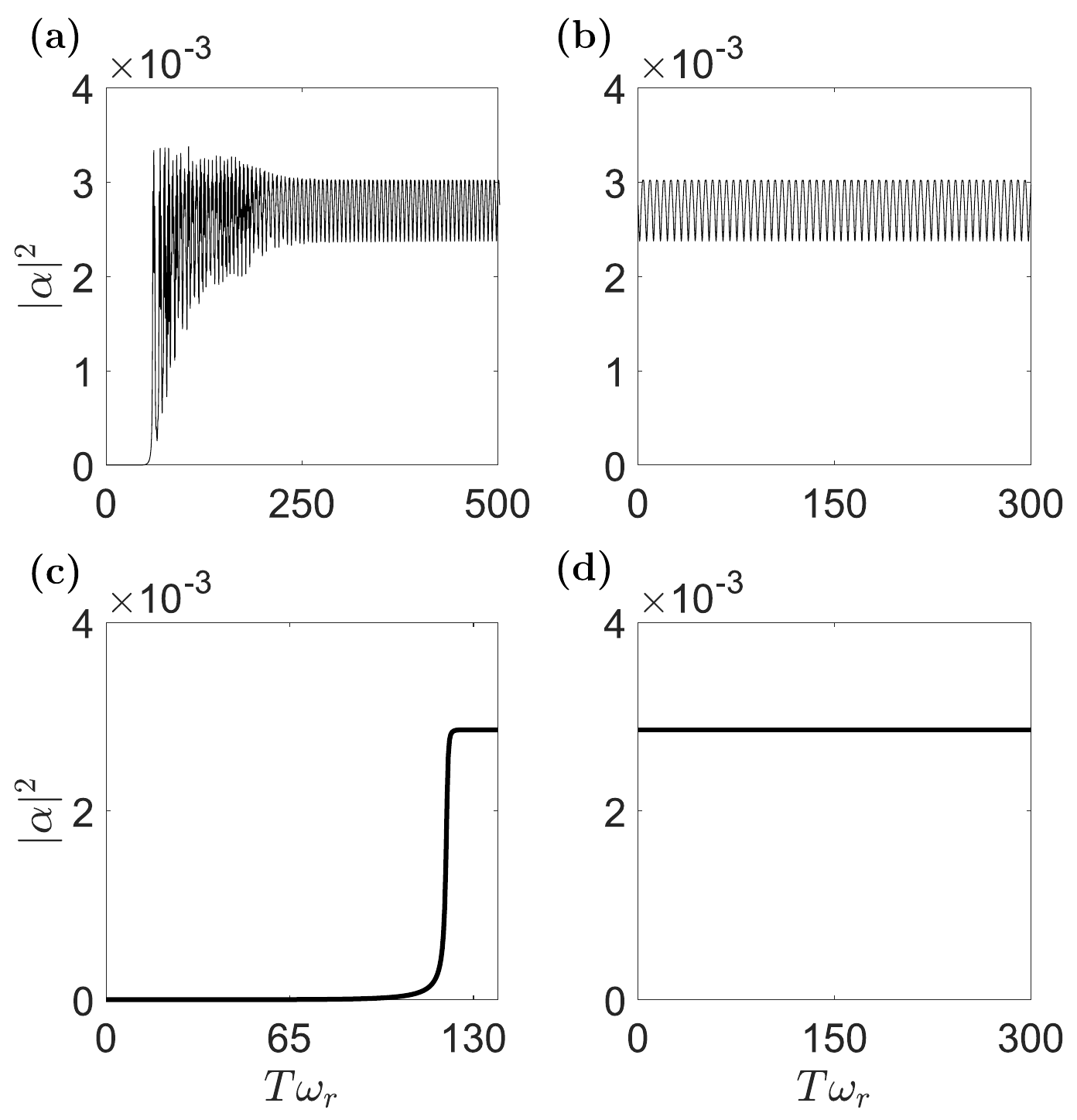}
		\caption{(a) Real time quench dynamics of the rescaled cavity field intensity for the limit cycle phase, where we simulate Eqs.~\eqref{eq3} and \eqref{eq:mfcavity}. (b) Real time dynamics with adiabatic cavity field Eq.~\eqref{eq5}, where we initialized (b) with the final point of (a). (c) Imaginary time dynamics using Eq.~\eqref{eq5} and Eq.~\eqref{eq6}. (d) Real time dynamics where we initialized (d) with the final point of (c). The parameters are $\Delta_c = -1900\omega_r$ and $\sqrt{N}\eta = 22\omega_r$, while the other parameters are the same as in Fig.~\ref{fig3}}
		\label{fig6}
	\end{figure}

To answer this question we first simulate the full dynamics of the atoms and cavity field as observed in Fig.~\ref{fig6}(a) for a parameter point in the limit cycle phase. After we have reached the limit cycle after long times we continue the simulation with only the adiabatic time evolution using Eq.~\eqref{eq5}. The result is visible in Fig.~\ref{fig6}(b). We find a very clean oscillation which highlights that the existence of a limit cycle does not require the non-adiabatic corrections. Instead, only the relaxation dynamics to the limit cycle requires the non-adiabatic corrections. In that context the limit cycle can truly be seen as a stable orbit described by the non-linear mean-field Hamiltonian~\eqref{eq:mfhelcav}. In the purely adiabatic theory this stable orbit exists and is metastable just like the OS phase. To demonstrate the latter we have prepared the OS phase for the same point using the imaginary time evolution and is visible in Fig.~\ref{fig6}(c). After this we have evolved the stationary state of the imaginary time evolution with the adiabatic equations of motion in real time. The result in Fig.~\ref{fig6}(d) shows no changes in the photon field, highlighting that this configuration is also metastable. 

While both the limit cycle and OS phase are metastable states of the adiabatic equations of motion, adding non-adiabatic corrections will directly destabilize the OS phase and make the limit cycle an attractor. This destabilization occurs for long timescales at a rather slow rate [$\mathrm{Im}(\omega_{\mathrm{crit}})\sim10^{-3}\omega_r$ in Fig.~\ref{fig4}(a)]. Meaning it exhibits a lifetime which is very long compared to the typical dynamics of the atoms. For an experimental realization this can imply that although the true stationary state is a limit cycle the OS remains stable over experimentally accessible timescales.

Comparing the results of the phase diagram in Figs.~\ref{fig3} and \ref{fig4} with the experiment in Ref.~\cite{Baumann2010-uv} we find qualitative agreement although we use a simplified one-dimensional model, while the experimental system is effectively two-dimensional. In addition, in Ref.~\cite{Baumann2010-uv} they used a different protocol where the drive amplitude is slowly ramped up from an initial to a final value $\sqrt{N}\eta_f$. Instead, the dynamics we consider here would correspond to preparing the system in the steady state of the initial drive amplitude and then suddenly quench its value to $\sqrt{N}\eta_f$. This indeed results in a qualitative difference for large detunings $\Delta_c\gg NU_0$ where a sudden quench can inject significant excitations while a slow ramp leaves the dynamics close to the stationary state. Instead, at optomechanical resonance, $\Delta_c\sim NU_0$, both protocols will inject large excitations and the dynamical behavior becomes similar again. This might explain why the dynamical instabilities found in Ref.~\cite{Baumann2010-uv} match our findings qualitatively.

    \section{Conclusion \label{Sec:4}}
    In conclusion, we have studied the emergence of dynamical phases in the bad cavity limit. Dynamical phases in this atom-cavity system emerge due to a strong dependence of the cavity detuning on the atomic pattern that can shift the cavity onto an optomechanical resonance. Using perturbation theory in the atom-cavity coupling we derived a mean-field Hamiltonian for the atomic variables that includes the effects of the density-dependent dynamical Stark shift. At different orders of the perturbative expansion we distinguish between dynamical instabilities that solely require adiabatic terms and instabilities that only exist when one includes non-adiabatic corrections. In the former we mostly find chaotic dynamics with wild dynamical oscillations of the field intensity. In the regime where non-adiabatic corrections destabilize the stationary state we can find limit cycle phases that have a well defined oscillation frequency. Remarkably, these phases are metastable states of the adiabatic model. The limit cycle phase appears to be the unique long-time limit if one treats the cavity and atomic degrees on equal timescales. 

    For the analysis in this work we calculated the excitation spectrum of the unorganized and self-organized states. In the future it would be interesting to explore excitations of the limit cycle solutions. Our work shows that limit cycles solutions can be described by the mean-field Hamiltonian. Thus, we might expect that coherent excitations can already be studied from this simple non-linear Hamiltonian. Including retardation effects, on the other hand, would also allow one to study the relaxation into the limit cycle phases. This is interesting as one could use this insight to stabilize or destabilize coherent oscillations.

    This work sets the basis for modelling the behavior of experimental platforms implementing optomechanical resonances for sensing \cite{Gupta:2007,Purdy:2010} and of strongly-correlated atoms confined in optical resonators \cite{Larson:2008,Cormick:2012} by providing relatively simple ansatzs for analyzing the dynamics.

    \section*{Acknowledgments}
    This work was funded by the Deutsche Forschungsgemeinschaft (DFG, German Research Foundation) – Project-ID 429529648 – TRR 306 QuCoLiMa (“Quantum Cooperativity of Light and Matter’’), the DFG Forschergruppe WEAVE "Quantum many-body dynamics of matter and light in cavity QED" - Project ID 525057097, and the QuantERA project "QNet: Quantum transport, metastability, and neuromorphic applications in Quantum Networks" - Project ID 532771420. S.B.J. acknowledge support from Research Centers of the Deutsche Forschungsgemeinschaft (DFG, German Research Foundation): Project No. 277625399-TRR 185 OSCAR (A4,A5)."

    \bibliography{references.bib}

    \appendix

    \section{Derivation of non-adiabatic corrections\label{App:A}}
    In this section we show additional steps which we used to calculate the eliminated cavity field with non-adiabatic corrections given by Eq.~\eqref{eq23}. Starting with formally integrating Eq.~\eqref{eq:mfcavity} we get
    \begin{equation}
        \begin{aligned}\label{eq25}
            \alpha(t) \approx  -i\sqrt{N}\eta\int_0^t e^{{\int_0^{\Tilde{\tau}}}(i[\Delta_c - NU_0B(t - \Tilde{\tau}')] - \kappa)d\Tilde{\tau}'}\Theta(t - \Tilde{\tau})d\Tilde{\tau}.
        \end{aligned}
    \end{equation}
    Then by Taylor expanding the time-dependent collective variables to first-order, followed by carrying out the integration in the exponential we obtain
    \begin{widetext}
        \begin{equation}\label{eq26}
            \begin{aligned}
                \alpha(t) &\approx  -i\sqrt{N}\eta\int_0^t \exp{{\int_0^{\Tilde{\tau}}}\Big(i\big(\Delta_c - NU_0\big[{B}(t) - \Tilde{\tau}^{'}\dot{B}(t)\big]\big) - \kappa\Big)d\Tilde{\tau}'}\Big[\Theta(t) - \Tilde{\tau}\Dot{\Theta}(t)\Big]d\Tilde{\tau}, \\
                &\approx -i\sqrt{N}\eta\Theta(t) \int_0^t \exp{\Big(\big[i\big(\Delta_c - NU_0{B}(t)\big) - \kappa\big]\Tilde{\tau} + \frac{iNU_0}{2}\Dot{B}(t)\Tilde{\tau}^2\Big)}d\Tilde{\tau}
                \\& \quad + i\sqrt{N}\eta\Dot{\Theta}(t) \int_0^t \exp{\Big(\big[i\big(\Delta_c - NU_0{B}(t)\big) - \kappa\big]\Tilde{\tau} + \frac{iNU_0}{2}\Dot{B}(t)\Tilde{\tau}^2\Big)}\Tilde{\tau}d\Tilde{\tau}.
            \end{aligned}
        \end{equation}
    \end{widetext}
    We now recast Eq.~\eqref{eq26} in a more compact form
    \begin{equation}
        \begin{aligned}\label{eq27}
            \alpha(t) &\approx \mathcal{W}\int_0^{\infty}\big(1 + \mathcal{Y}\Tilde{\tau}^2\big)e^{-\mathcal{X}\Tilde{\tau}}d\Tilde{\tau} + \mathcal{Z}\int_0^\infty e^{-\mathcal{X}\Tilde{\tau}}\Tilde{\tau}d\Tilde{\tau},
        \end{aligned}
    \end{equation}
    where
    \begin{equation}
        \begin{aligned}
            &\mathcal{W} = -i\sqrt{N}\eta\Theta, \quad \mathcal{X} = -(i\delta_c - \kappa), \\& \mathcal{Y} = \frac{iNU_0}{2}\Dot{B}, \quad \mathcal{Z} = i\sqrt{N}\eta\dot{\Theta},
        \end{aligned}
    \end{equation}
  Note that we have put the upper bound of the integral to infinity assuming that $\exp(-\mathcal{X}t)\approx 0$ and have neglected the term of order $\Tilde{\tau}^3$ under the assumption that $\Delta_c, \kappa \gg \omega_r, \sqrt{N}\eta$. Also, the first exponential in Eq.~\eqref{eq27} has been Taylor expanded to first-order to obtain an expression that is consistent with our perturbation theory. Now solving the integrals we find
  \begin{equation}
        \begin{aligned}
            \alpha(t) &\approx \frac{\mathcal{W}}{\mathcal{X}}+\frac{\mathcal{Z}}{\mathcal{X}^2}+\frac{2\mathcal{W}\mathcal{Y}}{\mathcal{X}^3},
        \end{aligned}
    \end{equation}
    which results in Eq.~\eqref{eq23}.
\end{document}